\documentclass[sigconf,hyphens,table]{acmart}

\setcopyright{none} % No copyright notice required for submissions
\acmConference[Qualification exam Yuchen Liu]{ACM template}{Due 5 Feb 2021}{Indiana, USA}
\acmYear{2021}

\usepackage{verbatim}
\usepackage{balance}
\usepackage{epsfig,endnotes}
\usepackage{graphicx, caption, subcaption}
\PassOptionsToPackage{hyphens}{url}
\usepackage{hyperref}

\usepackage[toc,page]{appendix}
  
\usepackage{tikz}
\usepackage{amsmath}
\usepackage{color}
\usepackage{soul}
\usepackage{comment}
\usepackage[hyphens]{url}
\usepackage{multirow}
\usepackage[final,inline,nomargin]{fixme}

\begin{document}
\title{Privacy-preserving and Privacy-attacking Approaches for Speech and Audio - A Survey} % TODO: replace with your title
\author{Yuchen Liu$^{1}$, Apu Kapadia$^{1}$, Donald Williamson$^{2}$}

\address{
  $^1$ Indiana University Bloomington \\
  $^2$ The Ohio State University
  }

\begin{abstract}
In contemporary society, voice-controlled devices, such as smartphones and home assistants, have become pervasive due to their advanced capabilities and functionality. The always-on nature of their microphones offers users the convenience of readily accessing these devices. However, recent research and events have revealed that such voice-controlled devices are prone to various forms of malicious attacks, hence making it a growing concern for both users and researchers to safeguard against such attacks. Despite the numerous studies that have investigated adversarial attacks and privacy preservation for images, a conclusive study of this nature has not been conducted for the audio domain. Therefore, this paper aims to examine existing approaches for privacy-preserving and privacy-attacking strategies for audio and speech. To achieve this goal, we classify the attack and defense scenarios into several categories and provide detailed analysis of each approach. We also interpret the dissimilarities between the various approaches, highlight their contributions, and examine their limitations. Our investigation reveals that voice-controlled devices based on neural networks are inherently susceptible to specific types of attacks. Although it is possible to enhance the robustness of such models to certain forms of attack, more sophisticated approaches are required to comprehensively safeguard user privacy.

\end{abstract}

 % TODO: replace with your keywords

\maketitle

%False positives for these triggers can cause serious security problems,such as opening garage doors without permission, leakage of bank account information, or execution of unauthorized commands. 

\section{Introduction}
In recent years, the prevalence of voice-controlled devices in users' homes has increased significantly. As of 2019, there were over 3 billion voice assistants in use\footnote{\url{https://voicebot.ai/2019/12/31/the-decade-of-voice-assistant-revolution/}}, and this number is projected to grow to 8.4 billion by 2024. Many Internet-of-Things (IoT) systems now utilize voice control, including products such as Google Home, Amazon Echo, and Apple HomePod, which allow users to control their smart home devices using voice commands. With these devices, users can perform tasks like making online purchases, unlocking doors, adjusting room temperature, and modifying home security systems, all through spoken interactions. Smartphones are also ubiquitous, with over 80\% of the world's population owning one. As of today, there are approximately 6.37 billion smartphones in use worldwide\footnote{\url{https://www.bankmycell.com/blog/how-many-phones-are-in-the-world}}, and many people also use other voice-enabled devices such as smartwatches and earphones. In 2020, over 444.7 million wearable units were shipped globally\footnote{\url{https://www.statista.com/statistics/437871/wearables-worldwide-shipments/}}, many of which include voice-assistive technology like Apple Siri, Microsoft Cortana, and Google Now.

The proliferation of voice-controlled devices has enabled users to interact with them using voice commands, which has improved their convenience and accessibility. However, concerns about security and privacy have emerged, with more than 41\% of voice assistant users expressing privacy concerns, according to recent research from Microsoft\footnote{\url{https://https://techcrunch.com/2019/04/24/41-of-voice-assistant-users-have-concerns-about-trust-and-privacy-report-finds/}}. Malicious attacks \cite{alepis2017monkey,diao2014your} on voice-controlled devices can lead to serious security risks, including unlocking a person's home, stealing their credit card information, or triggering unintended purchases. For example, a 2017 Burger King commercial\footnote{\url{https://www.bbc.com/news/technology-39589013}} inadvertently triggered Google Home smart speakers and Android phones to read out Wikipedia information about its products, by saying "OK, Google. What is the Whopper burger?" in the ad.  Subsequently, an attacker edited Wikipedia to describe the Whopper as the "worst hamburger product" and another added cyanide to the list of ingredients. Although this attack was relatively harmless, false positives of this nature can have serious consequences. In another instance, a TV anchor's joke \footnote{\url{https://www.cbsnews.com/news/tv-news-anchors-report-accidentally-sets-off-viewers-amazons-echo-dots/}} caused Amazon Echo devices in users' homes to try to order a \$160 dollhouse, which raised concerns about unintended purchases. These incidents demonstrate the need for greater security measures to safeguard voice-controlled devices against malicious attacks.

Research into privacy-attacking and privacy-defending methods has been ongoing for many years in various domains other than audio. For instance, in the image domain, a survey by Serban \textit{et al.} \cite{serban2020adversarial} demonstrates how attackers can produce adversarial examples to deceive object recognition models and cause them to misclassify objects in an image. Additionally, it outlines how defenders can protect the system from such attacks or redesign the model. In the natural language processing (NLP) domain, Zhang \textit{et al.} \cite{zhang2020adversarial} conduct a survey of attacks and defenses on text data. Melis \textit{et al.} \cite{melis2017deep} describe how to mislead the iCub robot vision system with adversarial examples that trick the classification process. This study also proposes a defense mechanism to enhance the model's security against adversarial examples. More recently, the paper from Li \textit{et al.} \cite{li2023multi} investigates privacy threats posed by large language models, such as OpenAI's ChatGPT, within application-integrated APIs like New Bing. The work highlights the potential for more severe privacy risks than previously seen, and supports these claims with extensive experimentation and discussion on privacy implications.

With the rise in the number of voice-controlled device users, safeguarding audio privacy has become a major concern. It is important that users, researchers and the general public are aware of the attack and defense mechanisms that exist today. Therefore, this paper presents a comprehensive survey of recent techniques that aim to protect or attack speech and audio privacy. Our focus is on four main attack categories, including impersonation attacks, operating system attacks, ultrasonic attacks, and adversarial attacks. In addition, we classify defense mechanisms as either detect-only or complete defense. Notably, voice-controlled system stages are susceptible to different types of attacks and defenses.

The subsequent parts of this paper are structured as follows: Firstly, we will provide an overview of the threat model and taxonomy, and define important terminology related to privacy-protecting and privacy-attacking strategies. Next, we will delve into each attack and defense mechanism, discussing them in detail. Lastly, we will conclude the paper with a discussion on the topic and suggest potential directions for future research.

\section{The Threat Model}
\begin{figure}[tbp]
        \center{\includegraphics[scale=0.22]
        {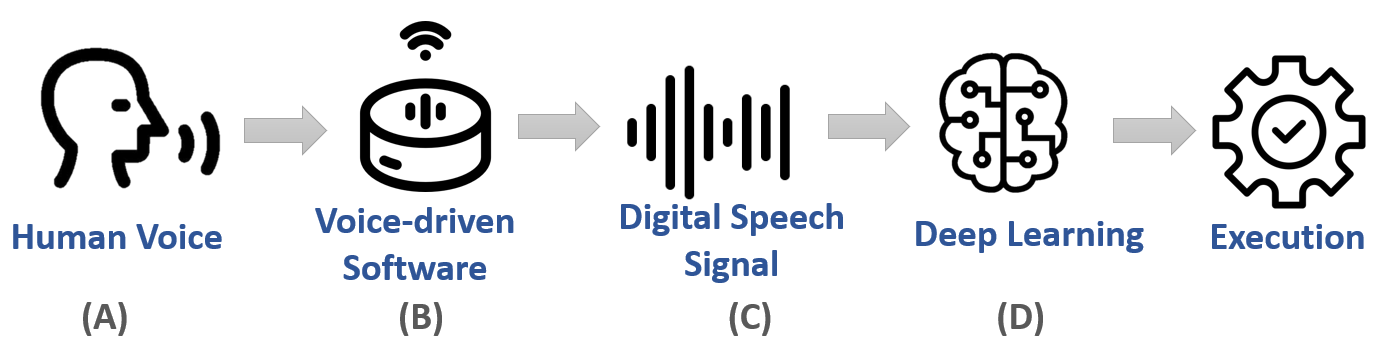}}
        \caption{The flow path of a typically voice-controlled system. 
        %Attacks can be placed at each stage: a) Spoofing the system using speech impersonations. b) Hacking into the operating system to force the voice-driven software to accept commands. c) Generating an ultrasonic signal that hides malicious commands. d) Using carefully crafted speech adversarial examples to fool the deep learning model. 
        }
        \label{fig:vcs-flow}
      \end{figure}

The basic flow of a voice-controlled system is illustrated in Fig. \ref{fig:vcs-flow}. Speech is first captured using a microphone and then converted into a digital signal before being provided to a deep learning model. The model performs automatic speech recognition and translates the signal into a computer-readable command, which is then executed. The attack and defense strategies can be implemented at any point in this process, and we have categorized them accordingly. 

We first define the threat model that describes the goal of the attacker and defender. We assume malicious attackers are interested in attacking a voice-controlled device of a target user. The attacker can target the device at any stage during the user's interaction, from when the user starts speaking to when the device executes the command. The goal of the attacker, also known as the adversary, is to confuse the original voice-controlled system and make the system to execute a malicious command without the target user noticing. The defenders are aware of possible attacks, so they try to either detect the incoming attack or reinforce the system to disable the attack. The basic assumption of attackers and defenders can be refined in several categories as follows:
\begin{description}
  \item[$\bullet$] \textbf{Attacker's Knowledge}: This category specifies how much an attacker knows about the system. \textit{White-box attacks} assume the attacker has full knowledge of the target system, including its setup and parameters. Attackers can replicate the model setup and parameters. In contrast, \textit{black-box attacks} assume that the attacker does not possess any information about the system, and must develop a general attack that can affect all types of systems.

  \item[$\bullet$] \textbf{Attacker's Goal}: The goal of the attacker depends on the type of attack. A \textit{targeted attack}, also known as a source-targeted attack \cite{papernot2016limitations}, aims to misclassify the input into a specific label or category. For instance, in the context of speaker verification, the attacker may want the system to recognize their voice as belonging to the target user. An \textit{untargeted attack}, on the other hand, does not require a specific label output. The adversary's aim is to misclassify the input into any incorrect class.
  
  \item[$\bullet$] \textbf{Physical/Logical Attacks}: A \textit{logical attack} involves the attacker injecting perturbations into the input speech in a simulated manner, such as through additive manipulation in software. This type of attack poses a limited real-world threat as the perturbation cannot play over-the-air. In contrast, a \textit{physical attack} occurs away from the device and allows the perturbation to play over-the-air, which means that these attackers must consider room acoustics.
  
  \item[$\bullet$] \textbf{Input specific/Universal Attacks}: An \textit{input specific attack} is dependent on the audio and targets each audio input specifically. Recently, \textit{universal attacks} have emerged \cite{moosavi2017universal}, whereby a single attack or perturbation can be applied to all inputs to the voice-controlled system. These attacks are potent because the attacker does not need any prior information about what the user is saying, and the attack occurs in real-time.
  
  \item[$\bullet$] \textbf{Defender's Goal}: From the defender's perspective, defense mechanisms can be classified into detection defenses or complete defenses. A \textit{detection defense} entails developing a classifier that detects whether the input has been modified or not, alerting the user if the input has been modified. On the other hand, a \textit{complete defense} not only detects the attack but also disables the attack by reducing its effectiveness.
\end{description}

%---------------------------
%% %---------------------------

\begin{table*}[]
\centering
\begin{tabular}{l|l|l}
\hline
Category                           & Type                     & paper                                                                        \\ \hline
\multirow{4}{*}{Privacy-attacking} & Impersonation Attack            & 

\begin{tabular}[c]{@{}l@{}} \cite{lindberg1999vulnerability}\cite{hunt1996unit}\cite{black1997automatically}\cite{tokuda2000speech}\cite{zen2009statistical}\cite{sebastien2014handbook} \cite{chen2017you}\cite{gibiansky2017deep}\cite{arik2017deep}\cite{ping2017deep}\cite{wang2017tacotron}\cite{sotelo2017char2wav}\cite{taigman2017voiceloop}\cite{mehri2016samplernn}\cite{toda2007one}\cite{kain2001design} \cite{toth2007using}\cite{toda2004acoustic}\cite{bearman2017accent}\cite{felps2009foreign}
                                   \\
\cite{hsu2016voice}\cite{biadsy2019parrotron}\cite{jia2018transfer}\cite{mohammadi2017overview} \cite{chen2021real}\cite{shen2018natural}\cite{faundez2006speaker}
                                   \end{tabular}  
\\ \cline{2-3} 
                                   & Operating System Attack  & \cite{jang2014a11y}
                                   \cite{diao2014your}
                                   \\ \cline{2-3} 
                                   & Ultrasonic Attack        & \cite{roy2018inaudible} \cite{zhang2017dolphinattack} \cite{song2017poster}  \\ \cline{2-3} 
                                   & Adversarial Example     & \begin{tabular}[c]{@{}l@{}} \cite{yuan2018commandersong} \cite{qin2019imperceptible} \cite{carlini2018audio} \cite{liu2020weighted} \cite{taori2019targeted} \cite{khare2018adversarial} \cite{schonherr2018adversarial} \cite{chen2020devil} \cite{neekhara2019universal} \cite{schonherr2020imperio} 
                                   \cite{yakura2018robust} \cite{li2019adversarial} \cite{ishida2020adjust} \cite{alparslan2020adversarial} \cite{dorr2020towards} \cite{alzantot2018did} \cite{abdullah2019hear}  
                                   \cite{latif2018adversarial}
                                   \\
                                   \cite{abdullah2019practical} \cite{guan2019surprising} \cite{vadillo2019universal} \cite{takahashi2020adversarial} \cite{li2020audio} \cite{ma2019detecting} \cite{vaidya2015cocaine} \cite{scardapane2017use} \cite{kreuk2018fooling} \cite{gong2017crafting} \cite{neekhara2019universal}
                            \cite{xu2018hasp}
                            \cite{gong2019real}
                         \cite{yang2020characterizing}
                         \cite{kurzinger2020audio}  
                         \cite{abdullah2021hear}
                         \cite{li2020advpulse}
                         \cite{zheng2021black}
                         \cite{guo2022specpatch}
                                   \end{tabular}                                                          \\ \hline
\multirow{2}{*}{Privacy-defending} 
                                   & Detection Only        &\begin{tabular}[c]{@{}l@{}} 
                                   \cite{kinnunen2017asvspoof}
                                  \cite{wu2015asvspoof}
                                  \cite{todisco2019asvspoof}\cite{lai2019assert}\cite{chettri2019ensemble}\cite{cai2019dku}\cite{bialobrzeski2019robust}\cite{lavrentyeva2019stc}\cite{yang2019sjtu}\cite{alluri2019iiit}\cite{li2019anti}\cite{williams2019speech}\cite{das2019long}\cite{chang2019transfer}\cite{gomez2019light}\cite{zeinali2019detecting}\cite{alzantot2019deep}\cite{jung2019replay}\cite{kamble2020advances}\cite{jelil2017spoof}
                                  \\              \cite{roy2018inaudible} \cite{lei2017insecurity}\cite{witkowski2017audio}\cite{todisco2017constant}\cite{bakar2018replay}\cite{lavrentyeva2017audio}\cite{li2017study}\cite{cai2017countermeasures}\cite{wang2017feature}\cite{chen2017resnet}\cite{jelil2018exploration}\cite{kamble2018effectiveness}\cite{yang2018feature}\cite{gong2018protecting}\cite{chen2017you}\cite{samizade2020adversarial}\cite{daubener2020detecting}\cite{rajaratnam2018isolated}\\ \cite{rajaratnam2018noise} \cite{jayashankar2020detecting} \cite{feng2017continuous} \cite{hussain2021waveguard} \cite{zhang2017dolphinattack}\end{tabular}  
                                                        \\ \cline{2-3} 
                                   & Complete Defense     & \cite{yang2018characterizing}\cite{mendes2020defending}\cite{das2018adagio}\cite{kwon2019poster}\cite{sun2018training}\cite{kurzinger2020audio}\cite{andronic2020mp3}\cite{esmaeilpour2020class} \cite{eisenhofer2021dompteur} \cite{yang2018towards}\cite{petracca2015audroid}\cite{esmaeilpour2019robust} \cite{zhang2017dolphinattack}  \cite{chen2020wearable} \cite{liu2021defending}  
                                   \cite{liu2022preventing}\\ \hline
\end{tabular}
\caption{Paper list of privacy-attacking and privacy-defending.}
\label{tab:paperlist}
\end{table*}

\section{Categories of Attacks}

In this section, we will examine recent techniques used to compromise privacy. The attacks can occur at any stage depicted in Fig \ref{fig:vcs-flow}. For instance, during the human voice stage, individuals can deceive the system by impersonating someone else's speech. At the voice-driven software stage, hackers can breach the operating system and commandeer the software to accept their orders. During the signal-to-digital converter stage, attackers can employ an ultrasonic signal to conceal their malicious commands. Finally, in the deep learning stage, individuals can employ well-crafted speech adversarial examples to deceive the deep learning model. A comprehensive list of papers on this topic can be found in Table \ref{tab:paperlist}. We will also provide an interpretation of these algorithms and discuss the advantages and disadvantages of each technique. The performance metrics presented in this section are based on the findings outlined in the original publications.

At the initial stage of voice-controlled execution, an attacker may carry out an impersonation spoofing attack on the voice-controlled system. This type of attack can be executed by employing a replay system \cite{chen2017you}, a synthetic speech system, or a converted speech system \cite{villalba2010speaker, wu2015spoofing}. By doing so, the attacker can prompt the voice-controlled system to execute their desired command. We discuss these attacks in section \ref{sec:impersonation}. In the second stage, the operating system can be targeted, where attackers may take over the system and force it to execute incorrect commands from the speaker \cite{diao2014your, jang2014a11y}. Additionally, the attacker may also gain access to the microphone in this stage \cite{anand2018speechless}. We describe these attacks in section \ref{sec:operatingsystem}. More recent attacks focus on the third and fourth stages. In contrast to the previous two stages, malicious actions in stage three and four are more difficult to detect and resolve. In the third stage, attacks typically utilize the non-linearity characteristic of the speech signal \cite{zhang2017dolphinattack, song2017poster, roy2018inaudible}. See section \ref{sec:ultrasonic} for more information. %These attacks craft specific ultrasonic signals that hide malicious commands. These ultrasonic signals are inaudible to the user and they force the voice-controlled device to execute malicious commands from the attacker. 

 %Deep learning models have been widely used in many privacy sensitive tasks today. For example, in image domain, we have deep learning models for privacy applications such as face recognition \cite{zhou20183d}, self-driving cars \cite{badue2020self}, medical image analysis \cite{litjens2017survey} and motion detection \cite{moeslund2001survey}. In the audio domain, the number of deep learning applications has experienced tremendous growth in the past few years, especially in the areas of speech recognition \cite{povey2011kaldi, padmanabhan2015machine}, speech enhancement\cite{lu2013speech, pascual2017segan}, speech emotion detection \cite{lalitha2019enhanced} and speaker verification\cite{snyder2017deep, snyder2016deep}. There is also an increasing number of audio-visual applications \cite{zhang2016multimodal, feng2017audio}. 
The final stage in the system involves feeding the signal into a deep learning model. In 2013, Szegedy discovered that certain adversarial attacks \cite{szegedy2013intriguing}, which had previously been effective in other domains \cite{goodfellow2014explaining, papernot2016transferability, papernot2017practical}, could also be employed to target deep learning models. Based on these findings, targeted audio adversarial attacks \cite{carlini2018audio} have been developed, which are particularly potent because human listeners cannot distinguish between the real audio and the adversarial example. Goodfellow argues that these adversarial examples exist due to the excessive non-linearity present in deep learning models \cite{goodfellow2014explaining}. Please refer to section \ref{sec:adversarial} for more information on deep-learning-based attacks.

% \usepackage{graphicx}
\begin{table*}[]
\centering
%\resizebox{\textwidth}{!}{%
\begin{tabular}{l|lllll}
\hline
Attack Name            & Attack Type             & \begin{tabular}[c]{@{}l@{}}Attacker \\Knowledge\end{tabular} & Attacker Goal & \begin{tabular}[c]{@{}l@{}}Physical/Logical \\ Attack\end{tabular} & \begin{tabular}[c]{@{}l@{}}Attack \\ Generality\end{tabular} \\ \hline
Replayed speech\cite{kinnunen2017asvspoof}        & Impersonation Attack    & Black Box          & Targeted      & Physical/Logical                                                   & Specific                                                     \\ \hline
Synthetic speech\cite{wu2015asvspoof}       & Impersonation Attack    & Black Box          & Targeted      & Physical/Logical                                                   & Specific                                                     \\ \hline
Converted speech\cite{todisco2019asvspoof}       & Impersonation Attack    & Black Box          & Targeted      & Physical/Logical                                                   & Specific                                                     \\ \hline
A11y attack\cite{jang2014a11y}            & Operating System Attack & White Box          & Targeted      & Physical/Logical                                                   & Specific                                                     \\ \hline
GVS attack\cite{diao2014your}            & Operating System Attack & White Box          & Targeted      & Physical                                                           & Specific                                                     \\ \hline
Dophin attack\cite{zhang2017dolphinattack}          & Ultrasonic Attack        & White Box          & Targeted      & Physical                                                           & Specific                                                     \\ \hline
Inaudible attack\cite{roy2018inaudible}             & Ultrasonic Attack        & White Box          & Targeted      & Physical                                                           & Specific                                                     \\ \hline
C\&W attack \cite{carlini2018audio}           & Adversarial Example     & White Box          & Targeted      & Logical                                                            & Specific                                                     \\ \hline
CommanderSong attack\cite{yuan2018commandersong}   & Adversarial Example     & White Box          & Targeted      & Physical/Logical                                                   & Specific                                                     \\ \hline
Imperceptible attack\cite{qin2019imperceptible}   & Adversarial Example     & White Box          & Targeted      & Physical/Logical                                                   & Specific                                                     \\ \hline
Robust physical attack\cite{yakura2018robust} & Adversarial Example     & White Box          & Targeted      & Physical                                                           & Specific                                                     \\ \hline
Black box adversarial attack\cite{taori2019targeted} & Adversarial Example     & Black Box          & Targeted      & Logical                                                            & Specific                                                     \\ \hline
Unviersal audio attack\cite{neekhara2019universal} & Adversarial Example     & White Box          & Untargeted    & Logical                                                            & Universal                                                    \\ \hline
\end{tabular}%
%}
\caption{Threat model of representative privacy-attacking techniques.}
\label{tab:attack}
\end{table*}

\subsection{Impersonation Attacks}\label{sec:impersonation}
Impersonation attacks, also referred to as spoofing attacks \cite{todisco2019asvspoof}, are the most fundamental type of attack on a voice-controlled system. In such attacks, the attacker creates a voice command that resembles the voice of the user of the smart voice assistant. Impersonation attacks can be classified into three types: synthetic speech attacks, converted speech attacks, and replayed speech attacks.

\subsubsection{Replay attacks}
The replay speech attack is the most common form of attack. Attackers use recorded speech of the target user to mimic their voice. For instance, the attacker can easily download the user's voice from their social media page \footnote{\url{https://audioboom.com/}}. Alternatively, the attacker can create a spam call that tricks the target user into saying a particular word or phrase that they desire. They can then use the recording of this phrase to launch an attack on the voice-controlled system \footnote{\url{https://www.varnumlaw.com/newsroom-publications-recording-conversations-with-your-cellphone-with-great-power-comes-potential}}. To execute a replay speech attack, the attacker must obtain a large amount of speech data from the target user.

\subsubsection{Synthetic speech attacks} 
Synthetic speech attacks use text-to-speech synthesis (TTS) to create simulated human voice commands that sound as if they originated from the target user \cite{gibiansky2017deep, arik2017deep, ping2017deep, wang2017tacotron, sotelo2017char2wav, taigman2017voiceloop, mehri2016samplernn}. Traditional TTS techniques primarily focus on concatenative synthesis \cite{hunt1996unit,black1997automatically} and parametric speech synthesis \cite{tokuda2000speech,zen2009statistical}. Once the attacker obtains enough recordings, they can extract the victim's acoustic model \cite{lindberg1999vulnerability}. By using the acoustic model, the attacker can reconstruct any desired commands through speech synthesis techniques\footnote{\url{https://www.pro-tools-expert.com/home-page/2016/11/16/adobe-voco-should-we-be-afraid}}. However, the resulting audio clips often sound artificial and unnatural due to the noise and reverberation present in the recorded speech.

\begin{figure}[tbp]
        \center{\includegraphics[scale=0.37]
        {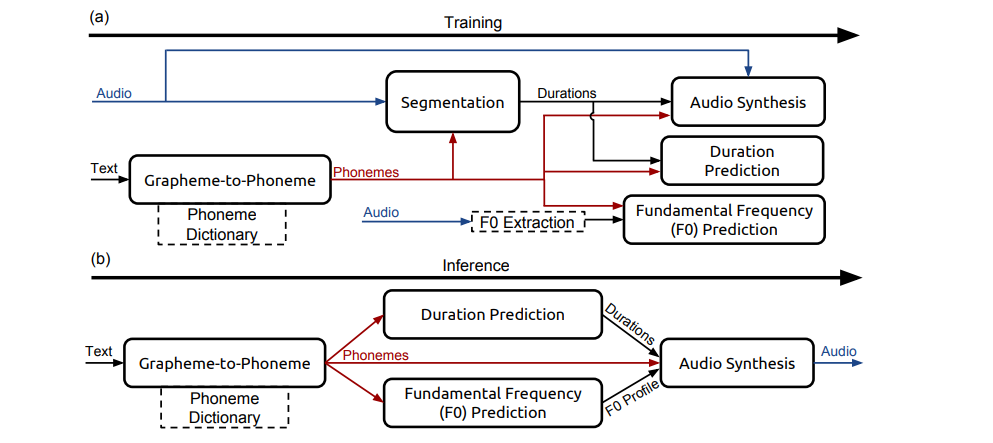}}
        \caption{Model Structure of Deep Voice from Baidu Inc. \cite{arik2017deep}}
        \label{fig:deepvoice}
      \end{figure}

Modern TTS methods use conventional source-filter vocoders \cite{morise2016world,kawahara2006straight} or a WaveNet-based vocoder \cite{oord2016wavenet} to produce more natural-sounding speech. A vocoder is an electronic device or software that is used to analyze and synthesize speech or other sounds. It works by breaking down the incoming sound signal into its spectral components and then re-synthesizing it using a carrier signal to produce an output that sounds like the original sound but with different characteristics. In Arık \textit{et al.} \cite{arik2017deep},  a real-time text-to-speech system called `Deep Voice' is proposed that uses a WaveNet-based vocoder.  The detailed model structure is shown in Fig \ref{fig:deepvoice}. In this system, the text data is first converted into phonetic information, which is then fed into a segmentation model that identifies where each phoneme begins and ends. The duration model predicts the duration of each phoneme, and the fundamental frequency module predicts whether a phoneme is voiced or not. The audio synthesis model combines the outputs from each module and the vocoder to generate a synthesized audio signal. This audio can be used for a synthetic speech attack. 

%---------------------------
%% %---------------------------

More recent approaches focus on end-to-end TTS conversion, as demonstrated in Ping \textit{et al.} \cite{ping2017deep}. End-to-end TTS uses an encoder to convert text into an internal latent representation and a decoder decodes this representation into an audio spectrogram.  A vocoder then transforms the predicted features into a speech signal. Similar architectures are also found in Tactron \cite{wang2017tacotron} and Tacotron 2 \cite{shen2018natural}. End-to-end TTS models differ from traditional methods, where with traditional approaches each module is trained separately and each component requires prior knowledge about the text to make speech synthesis possible.

The synthetic speech attack is considered to be a black-box attack. The attacker does not need any information or knowledge from the user. However, this type of attack may be disabled if the voice-controlled system has a speaker verification (SV) system that can detect if the speech is from the original user or not.

\subsubsection{Converted speech attack}
\begin{figure}[tbp]
        \center{\includegraphics[scale=0.42]
        {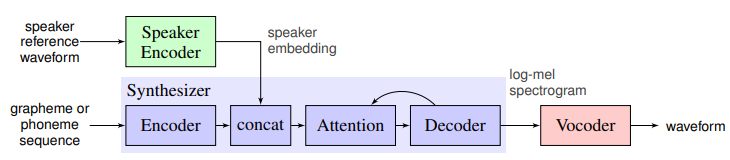}}
        \caption{Model Structure of Google Voice Conversion system \cite{jia2018transfer}}
        \label{fig:googlevc}
      \end{figure}

Converted speech attacks are similar to synthetic speech attacks in that they aim to generate speech that mimics the target user's voice. However, converted speech attacks differ in that they use voice conversion (VC) techniques to make the resulting speech sound like that of the target user \cite{toda2007one}. Many techniques for performing VC have been proposed in recent years. In multiple studies \cite{kain2001design,toth2007using,toda2004acoustic}, the authors use Gaussian-mixture models (GMMs) within the VC system to capture the statistical properties of the acoustic features for both source and target speakers. More recent approaches use neural networks \cite{hsu2016voice, biadsy2019parrotron}. An overview of a VC system can be found in Mohammadi \textit{et al.} \cite{mohammadi2017overview}. VC systems typically use an encoder-decoder architecture, similar to that used in text-to-speech systems. For example, Jia et al. \cite{jia2018transfer} proposed a VC system that generates speech for a specific speaker using only five seconds of their speech. Fig \ref{fig:googlevc} shows the model structure of the Google VC system. The system uses a separately trained speaker encoder to extract the speaker encoding from the speaker reference utterance, and then the encoder of the synthesizer extracts the speaker-independent information from the original speech to concatenate it with the speaker encoding data to form a speaker-dependent audio representation. The audio representation is then fed into the decoder to generate a log-mel spectrogram feature, which is transformed into an audio utterance by a vocoder, where the resulting audio sounds like the target user and it contains a malicious command.

%---------------------------
%% %---------------------------

Converted speech attacks are considered the strongest type of impersonation attack. Additionally, some VC systems can even simulate the accent of the speaker, making it more difficult to detect and defend against such attacks \cite{bearman2017accent, felps2009foreign}.

\subsection{Operating System Attacks} \label{sec:operatingsystem}
Operating system (OS) attacks, as the name implies, exploit vulnerabilities within the OS to execute attacks. These attacks are self-triggered and more difficult to detect compared to impersonation attacks. In this section, we discuss two notable OS attacks: \textit{A11y attacks} \cite{jang2014a11y} and \textit{Google Voice Search (GVS) attacks} \cite{diao2014your}.
 
\subsubsection{A11y attacks}
In 1998, the Rehabilitation Act of 1973\footnote{\url{https://www.section508.gov/Section-508-Of-The-Rehabilitation-Act}} was amended by the United States Congress with the objective of making it easier for individuals with disabilities to use electronic devices and information technology.  As a result, recent operating system developers have integrated various accessibility features, such as voice commands, speech recognizers, and on-screen keyboards, among others, into their OS. \footnote{\url{https://www.google.com/sites/accessibility.html}}\footnote{\url{https://www.microsoft.com/en-us/accessibility/}}\footnote{\url{https://www.apple.com/accessibility/}} However, these accessibility technologies have also brought security concerns. In Jang \textit{et al.} \cite{jang2014a11y}, the author introduced malware that can be used to exploit commonly-used operating systems. In total, Jang presented 12 different OS attack approaches that leverage different accessibility tools. This include Windows attacks: 1.~privilege escalation through Speech Recognition, 2.~privilege escalation with Explorer.exe, 3.~stealing passwords using Password Eye and a screenshot, 4.~stealing sudoer passwords from authentication dialogsm; Ubuntu Linux attacks: 5.~bypassing the security boundaries of Ubuntu; IOS attacks: 6.~bypassing passcode lock using Siri, 7.~bypassing the iOS sandbox, 8.~privilege escalation with remote view, 9.~bypassing password protection on iOS; and Android attacks: 10.~bypass Touchless Control’s voice authentication, 11.~bypassing Android sandboxing, and 12.~keylogger on Android.
In this section, we will discuss three speech- and audio-related approaches that are used in voice-based attacks.

\begin{figure}[tbp]
        \center{\includegraphics[scale=0.42]
        {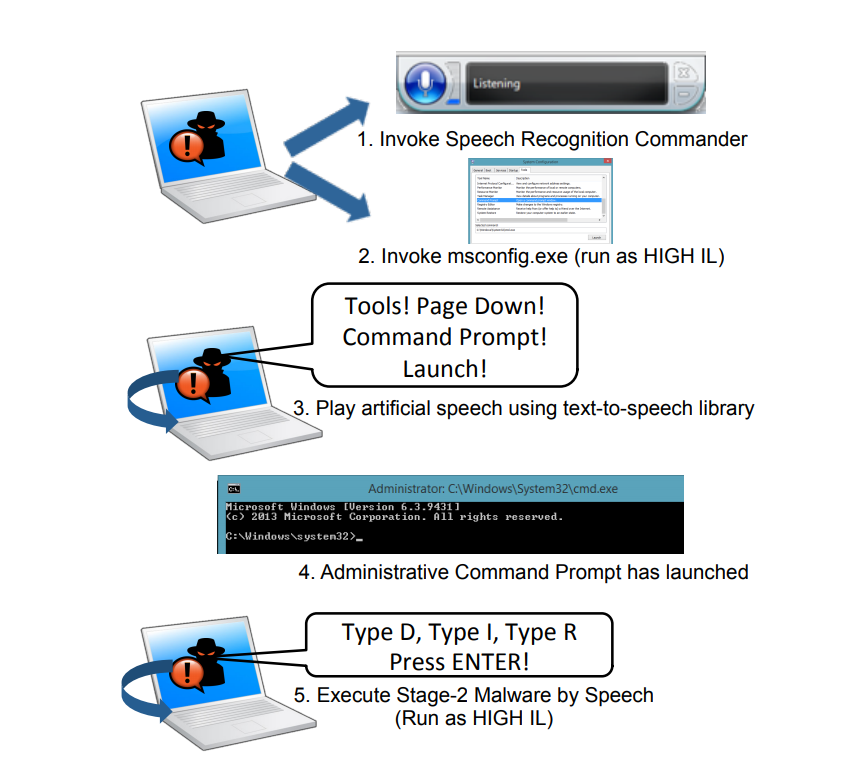}}
        \caption{Windows A11y attack using speech recognition commander accessibility tool \cite{jang2014a11y}}
        \label{fig:a11ywindows}
      \end{figure}

In the case of Windows OS, attackers can take advantage of the speech recognition accessibility feature to obtain higher privileges. The speech recognition system in Windows always grants administrative privileges at a high integrity level (High IL). The attack scenario is illustrated in Fig \ref{fig:a11ywindows}. Firstly, the attacker uses CreateProcess() with the argument sapisvr.exe -SpeechUX to launch the speech recognition accessibility tool. Then, the malware is employed to open the msconfig.exe application via CreateProcess(). By default, the application runs in High IL. Next, the attacker can issue a voice command with the transcript "Tools, Page down, Command prompt, Launch" to open the command shell in Windows. Finally, the opened command shell inherits the High IL, providing command line access with administrative privileges, which allows the attackers to execute any desired command.

%---------------------------
%% %---------------------------

Accessibility features in mobile operating systems, such as iOS and Android, can also be exploited for attacks. In the case of iOS, an attacker can use Siri to bypass the password lock screen and access user-sensitive data or perform security-related commands, even when the screen is locked \footnote{\url{https://www.businessinsider.com/password-security-flaw-in-ios-7-lets-siri-control-your-iphone-2013-9}}. iOS allows Siri to bring user-sensitive data or make security related commands even when the screen is locked. Therefore, this can be done without any knowledge of the user's password. Similarly, for Android devices, an attacker can use malware as a background service to record the user's voice constantly. When the phrase "OK Google Now" is detected, the attacker can perform a replay speech attack using the device's microphone. Following this, the attacker can use a synthetic speech attack and TTS speech commands to deceive the Google Now based voice-controlled system.

\subsubsection{Google Voice Search (GVS) attack}
In Diao \textit{et al.} \cite{diao2014your}, the author generated malware using a zero-permission Android application called VoicEmployer to attack the Android built-in voice assistant tool Google Voice Search. 

Based on Diao's research, Google voice search has two different modes: voice dialer mode and velvet mode. The author shows that a third-party app using the Bluetooth module can pass an ACTION\_VOICE\_COMMAND based intent to the Android operating system and trigger the voice dialer mode of Google Voice Search even though the device is locked.

%---------------------------
      \begin{figure}[tbp]
        \center{\includegraphics[scale=0.42]
        {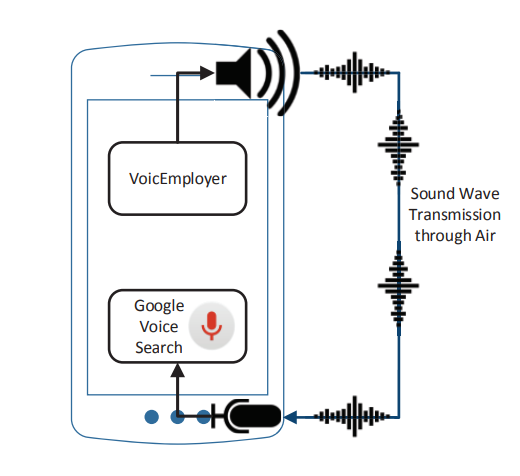}}
        \caption{GVS attack using VoicEmployer malware \cite{diao2014your}}
        \label{fig:GVSattack}
      \end{figure}
%% %---------------------------

Fig \ref{fig:GVSattack} shows the inter-application communication of the GVS attack. The malware VoicEmployer first uses the speaker in the device to produce the attack audio that activates Google voice search in the voice dialer mode. VoicEmployer then continuously analyzes the environment and places the attack, if the user is not nearby or sleeping. The malware then places a low sound impersonation attack lower than 55dB so that the user cannot notice the attack \cite{muzet2007environmental}.

\subsection{Ultrasonic Attack} \label{sec:ultrasonic}
 The upper bound frequency of human hearing is 20 kHz, where human speech often occurs at frequencies much lower than this. As a result, most voice-controlled systems use low-pass filters to remove signal components that occur above 20 kHz \cite{lee2015chirp}, since the spoken command will be contained in lower frequencies.  Unfortunately, attackers have developed sophisticated workarounds that enable them to generate commands in frequency ranges above 20 kHz.  In particular, ultrasonic sounds have frequencies higher than 20kHz, so they are inaudible to humans. With ultrasonic attacks, the attackers leverage the non-linearity of the microphone and speaker in the voice controlled system and use it to hide inaudible commands in the ultrasonic frequency range, so that the target device still receives the commands. These attacks can be really damaging as they are difficult to detect. Next, we introduce two representative ultrasonic attacks: \textit{dolphin attacks} \cite{zhang2017dolphinattack} and \textit{inaudible voice command attacks} \cite{roy2018inaudible}.

\subsubsection{Dolphin attacks}
\begin{figure}[tbp]
        \center{\includegraphics[scale=0.42]
        {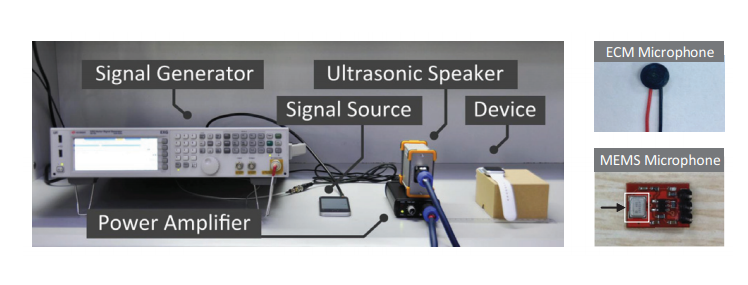}}
        \caption{Device setup for Dolphin attack \cite{zhang2017dolphinattack}}
        \label{fig:da1}
      \end{figure}

\begin{figure}[tbp]
        \center{\includegraphics[scale=0.42]
        {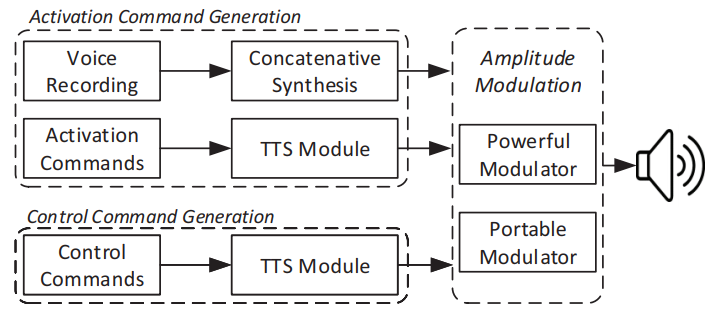}}
        \caption{Model Architecture for Dolphin attack \cite{zhang2017dolphinattack}}
        \label{fig:da2}
      \end{figure}

In Zhang \textit{et al.} \cite{zhang2017dolphinattack}, the authors exploit the non-linearity feature of a Micro Electro Mechanical System (MEMS) microphone and an Electret Condenser microphone (ECM) to generate inaudible ultrasonic signals that can carry malicious commands. 
The experiment device setup is shown in Fig. \ref{fig:da1} and the detailed model architecture for a Dolphin attack is shown in Fig. \ref{fig:da2}. The attack starts with voice command generation, which consists of two parts: activation command generation and general control command generation. Activation command generation refers to the process of generating a specific phrase or word that triggers the voice-controlled device to start listening for a user's command. General control commands refer to the voice commands given by the user to control the device's various functions. Since the activation command needs to pass a speaker verification process and the general control command does not, the attacker generates the activation command by using a concatenative synthesis technique that generates the wake words by concatenating different phonemes from other recordings from the target user similar to the impersonation attack we mentioned earlier. For example, the wake-up words `Hey Siri' can be generated from “\underline{he} is a boy”, “eat a c\underline{a}ke”, “in the \underline{ci}ty”, “\underline{re}ad after me”. The general control command can be simply generated by any state-of-art TTS system.

After the voice command is successfully generated, the author uses amplitude modulation (AM) to generate the ultrasonic signal. Amplitude modulation is a modulation technique used to transmit information through a carrier wave by varying the amplitude of the carrier wave in accordance with the information to be transmitted. An ultrasonic carrier is chosen based on the modulation depth $m$ and carrier frequency $f_{c}$. These two parameters are hardware dependent and they decide the amplitude and frequency of the final ultrasonic signal. After the ultrasonic signal has been generated, a powerful transmitter then transmits the signal to the target voice-controlled device. The result shows that the longest attack distance with an inaudible signal is about 175cm.

%---------------------------
%% %---------------------------

%---------------------------
%% %---------------------------

\subsubsection{Inaudible Voice Commands attack}
In the dolphin attack, the attack range has a limit of approximately 175cm. If we want to increase the attack range by using a more powerful ultrasonic transmitter, then audio leakage may occur that makes the signal audible to the target user, e.g., part of the ultrasonic signal is leaked into the human-audible frequency range. In Roy \textit{et al.} \cite{roy2018inaudible}, the author proposes an approach that can place a long-range ultrasonic attack, without ultrasonic leakage. This kind of attack is more dangerous than the previous one \cite{zhang2017dolphinattack} since the attack can occur from outside of a person's home. 

\begin{figure}[tbp]
        \center{\includegraphics[scale=0.20]
        {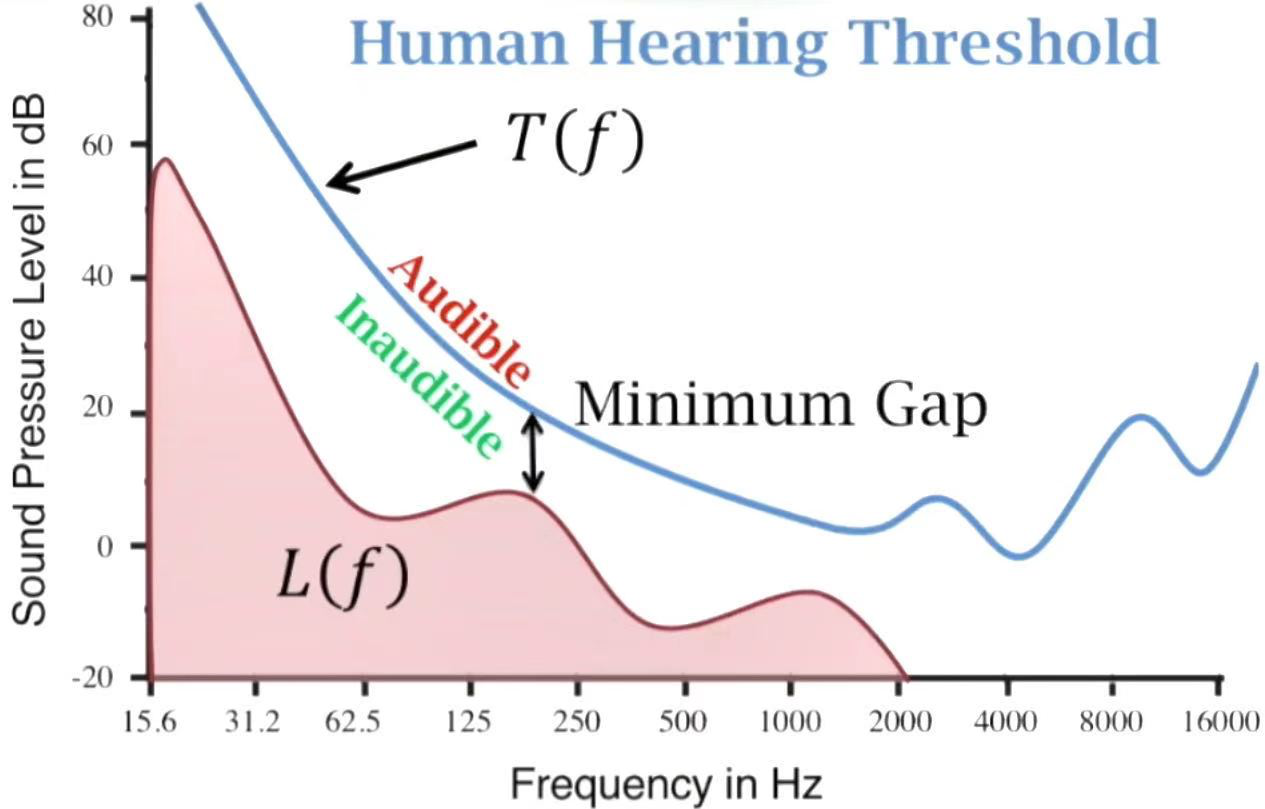}} 
        \caption{Main idea of Roy \textit{et al.} \cite{roy2018inaudible} to minimize the gap between the human hearing threshold and speaker leakage.}
        \label{fig:hearingthres}
      \end{figure}

\begin{figure}[tbp]
        \center{\includegraphics[scale=0.53]
        {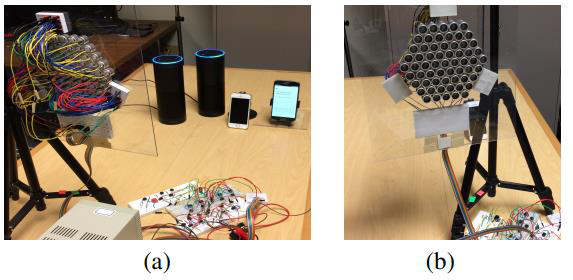}}
        \caption{(a) Device setup in Roy \textit{et al.} \cite{roy2018inaudible}. (b) The ultrasonic microphone array for the attack.}
        \label{fig:speakerarray}
      \end{figure}

To solve this problem, Roy \textit{et al.} found that not only the microphone has the non-linearity feature, but the loudspeaker also has similar characteristics. Therefore, the author utilizes the loudspeakers non-linearity feature and formulates this question as an optimization problem. This is done in order to hide the audible audio leakage spectrum, $L(f)$, into the human hearing threshold $T(f)$ as Fig. \ref{fig:hearingthres} shows, so that human listener will not hear the signal. The approach uses an ultrasonic loudspeaker array with 61 loudspeakers, as shown in Fig \ref{fig:speakerarray}. Each microphone helps to segment the input signal into a small ultrasonic signal piece. The result shows that the attack can be successfully placed when the loudspeaker is 12ft away from the device. The author also proposes several detection-based defense mechanisms, and these techniques are introduced in section \ref{sec:detect-only}. 

%---------------------------
%% %---------------------------

%---------------------------
%% %---------------------------

 Ultrasonic attacks are strong but they also have drawbacks. These attacks require specific ultrasonic transducers to produce the ultrasonic signal. Furthermore, for long-range attacks, a big and powerful ultrasonic speaker is needed, if you want to attack outside of the target user's home. Hence, the attacker needs to be close to the device in order for this attack to occur.

\subsection{Adversarial Attacks} \label{sec:adversarial}
\begin{figure}[tbp]
        \center{\includegraphics[scale=0.55]
        {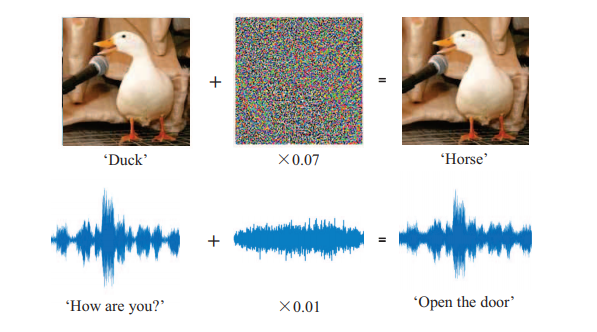}}
        \caption{Adversirial example in image and audio domain Gong \textit{et al.} \cite{gong2018overview}.}
        \label{fig:adversirialexmaple}
      \end{figure}

Modern voice-controlled systems are often equipped with a state-of-art automatic speech recognition (ASR) system,  such as Deep Speech \cite{hannun2014deep}, Lingvo \cite{shen2019lingvo}, Kaldi \cite{povey2011kaldi}, to name a few. These deep learning  based approaches have great performance on recognition tasks with about 5\% word error rate (WER). However, it has been shown that DNN models have vulnerabilities, as is discussed in Szegedy \textit{et al.} \cite{szegedy2013intriguing}. Fig. \ref{fig:adversirialexmaple} shows the basic idea of an adversarial attack in both the image and audio domains. This type of attacks aims to mis-classify the original label to a different label by adding a certain perturbation to the input. For images, this may result in the classifier incorrectly identifying the object in the image. For speech recognition, the resulting audio signal may sound (and look like) the original input, but the ASR system may transcribe the audio incorrectly, where it is transcribed as a malicious command of the attacker. 

%---------------------------
%% %---------------------------

Cocaine Noodles \cite{vaidya2015cocaine} and Hidden Voice Command \cite{carlini2016hidden} are the first two approaches that discuss the vulnerabilities associated with automatic speech recognition. They found that the ASR systems are highly reliant on acoustic features, such as Mel-frequency cepstral coefficients (MFCCs). Using this feature, they successfully generate adversarial examples that contain enough acoustic features that the ASR system accepts it. However, these attack approaches do have limitations. The output speech, after adversarial noise is added, is not understandable, so the user may notice the attack and take effective defense measures. Recent adversarial approaches, however, successfully produce utterances that are still fully understandable by humans, but that are mis-classified by state-of-art ASR systems. In the following subsections, we introduce different adversarial example approaches based on the threat model of attackers.

\subsubsection{C\&W attack}
In 2018, Carlini and Wagner \cite{carlini2018audio} first successfully placed an end-to-end targeted adversarial attack on the Deepspeech ASR system \cite{hannun2014deep}. They found that for any input $x$, it is possible to find a small $\delta$ to generate $x^{\prime}$ where $x^{\prime}=x+\delta$ so that $x$ and $x^{\prime}$ sound nearly identical. However, when $x^{\prime}$ is provided as the input to the ASR system, i.e., $f\left(x^{\prime}\right)$, the ASR outputs $y$, which is a malicious command. 

In order to make $x$ and $x^{\prime}$ sound similar, Carlini and Wagner use the decibel (dB) as a distortion metric. The  relative loudness of an audio sample is calculated as:
\begin{equation}
d B(x)=\max _{i} 20 \cdot \log _{10}\left(x_{i}\right)
\end{equation}
\noindent
The distortion level between the original waveform, $x$, and the added perturbation, $\delta$, can be calculated as:
\begin{equation}
d B_{x}(\delta)=d B(\delta)-d B(x)
\end{equation}
The problem now can be formulated as the following optimization problem:

\begin{equation}
\begin{array}{l}
\text { minimize } d B_{x}(\delta) \\
\text { such that } f(x+\delta)=t \\
\qquad x+\delta \in[-M, M]
\end{array}
\end{equation}
$M$ here means the maximum representable value for the adversarial example which can be accomplished by clipping. Here, $t$ is the malicious command transcript, which is the target label the attacker wants to achieve. Due to the non-linearity of the constraint
$f(x+\delta)=t$, the optimization problem requires an additional loss term:

\begin{equation}
\operatorname{minimize} d B_{x}(\delta)+c \cdot \ell(x+\delta, t)
\end{equation}
Here $\ell(\cdot)$ represents the additional loss term. Smaller values for $\ell(x+\delta, t)$ indicate that the predicted transcript is closer to target transcript. $c$ here helps to control the weights of the distribution level and adversarial performance. In this paper, the author uses the connectionist temporal classification (CTC) loss \cite{graves2006connectionist}. This is a commonly used loss function in speech recognition tasks. A lower CTC loss indicates that the output transcription is closer to the ground truth label. The final difficulty is that the system may fail to converge when the inserted perturbation is too small. Therefore, the author sets a constant $\tau$ and forces the system to converge when $d B_{x}(\delta) \leq \tau$. If the system successfully converges, $\tau$ can be reduced iterating until no solution can be found. Therefore, the final optimization problem can be described as:

\begin{equation}
\begin{array}{l}
\text { minimize }|\delta|_{2}^{2}+c \cdot \ell(x+\delta, t) \\
\text { such that } d B_{x}(\delta) \leq \tau
\end{array}
\end{equation}

By using the above method, the author reaches a 100\% success attack rate with a 99.9\% similar adversarial example compared to the original audio clip. Since this is a white box, non-universal, logical access threat model, C\&W attacks pose a limited threat to voice-controlled systems. %We next introduce more recent over-the-air attacks that are universal and black box attacks.

\subsubsection{Over-the-air attacks}
In this section, we introduce some representative audio adversarial over-the-air attacks. These attacks are more dangerous than logical attacks, because they can attack the target device physically and from a long distance. The representative publications are \cite{yuan2018commandersong,qin2019imperceptible,yakura2018robust}.

In Yuan \textit{et al.} \cite{yuan2018commandersong}, the author proposes a white box attack that generates adversarial music that contains a malicious command to attack a Kaldi ASR system \cite{povey2011kaldi}. The attack can be either logical (WAV-To-API, WTA), which is simulated digitally or physical (WAV-AIR-API, WAA), which can be deployed through the air. A Kaldi ASR system contains multiple components such as an acoustic model and a language model. The acoustic model in it can be trained with a DNN and it represents the probability between input features and phonemes. The language model then represents the probability distribution over the sequence of words. For WTA attacks, the author tried to use a probability density function index (pdf-id) sequence matching method to hide the command audio into the song audio. This method involves creating a targeted command by replacing certain phonetic units in the original command with other units that have similar acoustic features, but a different meaning.

In order to make the attack over-the-air and accomplish WAA attack, the author added a noise model to simulate the background noise and electronic noise of speakers to the pdf-id sequence matching model $x^{\prime}(t)=x(t)+\mu(t)$, where $x(t)$ is the result from a WTA attack and $\mu(t)$ is the random noise model. The author reports 100\% success rate on WTA attacks and achieves a 96\% success rate when using the JBL speaker with a 1.5m distance between the speaker and microphone. 

In Qin \textit{et al.} \cite{qin2019imperceptible}, the author improved the C\&W attack so that the attack can play over-the-air. The author proposed an attack scenario to attack the Lingvo ASR system \cite{shen2019lingvo}. The author made the perturbation further imperceptible by using frequency masking, where a softer sound becomes inaudible since it is obscured by a louder sound. The power spectral density, $p_{\delta}$, of the perturbation is calculated in each iteration to make sure it falls below the masking threshold, $\theta_{x}$, of the original utterance. 
The loss function is formulated as:
\begin{equation}
\ell(x, \delta, t)=\ell_{n e t}(f(x+\delta), t)+\alpha \cdot \ell_{\theta}(x, \delta)
\end{equation}
The first part of the equation is from the C\&W attack to make the audio produce the target label where $x$ is the original speech, $\delta$ is the added pertubation and $t$ is the target transcript. The second part of the equation controlled by weight $\alpha$ aims to make the perturbation imperceptible. The optimization problem can be then formulated as: 

\begin{equation}
\min _{\delta} \ell(x+\delta, t)+\alpha \sum_{k=0}^{\left\lfloor\frac{N}{2}\right\rfloor} \max \left\{p_{\delta}(k)-\theta_{x}(k), 0\right\}
\end{equation}
N here is the STFT window size. $p_{\delta}(k)$ is known as the power spectral density (PSD) of the perturbation and $\theta_{x}(k)$ is the frequency masking threshold of the original audio where $k$ represents the kth bin of the spectrum of frame x.

Besides the above contribution, the author also tries to make the attack physical. The author first simulates the room impulse, $r$, and convolves the speech with it to produce the reverberant signal, $C(x)=x * r, t \sim \mathrm{T}$. Then the loss function becomes:

\begin{equation}
 \ell(x, \delta, y)=\mathrm{E}_{t \sim \mathrm{T}}\left[\ell_{\text {net }}(f(C(x+\delta)), t)\right]+\alpha \cdot \ell_{\theta}(x, \delta)
\end{equation}
where the first part of the equation is the robustness loss and the second part of equation is for imperceptibility as before. In terms of results, the author first uses three experiments with Amazon Mechanical Turk users to evaluate the effectiveness of adversarial examples in audio manipulation, finding that users had difficulty distinguishing between clean and adversarial examples. These adversarial examples are sent to the Lingvo ASR system \cite{shen2019lingvo} and reach 49.65\% over-the-air accuracy and 22.98\% word error rate (WER) while keeping the perturbation imperceptible.

In Yakura \textit{et al.} \cite{yakura2018robust}, the author successfully placed an over-the-air attack to the Deepspeech ASR system. In order to aid the robustness of the audio adversarial example and make the over-the-air attack possible, the author introduces three techniques to simulate the transformations caused by playback and recording, into the generation process. The three components are band-pass filtering, room impulse response, and white Gaussian noise. The author started by using the original loss function from C\&W attack to generate the audio adversarial audio as follow:

\begin{equation}
\underset{\boldsymbol{v}}{\operatorname{argmin}}  \:{\ell} \underset{f} (M F C C(\boldsymbol{x}+\boldsymbol{\delta}), t)+\epsilon\|\boldsymbol{\delta}\|
\end{equation}
Here $x$ and $\delta$ represent the original speech signal and the added perturbation. $MFCC(\cdot)$ indicates the MFCC feature extraction from the mixed signal $x+\delta$. After the logical audio adversarial example is successfully generated, the author adds robustness to the function. First, the author uses a band-pass filter to limit the frequency range of the perturbation. As was introduced before, modern microphones are often made to automatically cut off the inaudible range of the signal. Therefore, the author limits the frequency bands to 1k to 4k Hz. The loss function is updated as follow:

\begin{equation}
\begin{array}{l}
\underset{\boldsymbol{\delta}}{\operatorname{argmin}} \: \ell(M F C C(\tilde{\boldsymbol{x}}),t)+\epsilon\|\boldsymbol{\delta}\| \\
\text { where } \tilde{\boldsymbol{x}}=\boldsymbol{x}+\underset{1k \sim 4k \mathrm{~Hz}}{B P F}(\boldsymbol{\delta})
\end{array}
\end{equation}
An impulse response represents the reaction obtained when an audio system is presented with a brief input signal, called an impulse. In the second step, the author aims to make the generated adversarial examples robust against reverberation by incorporating impulse responses from various environments into the generation process \cite{peddinti2015reverberation}. Similar as \cite{athalye2018synthesizing}, the author comuptes the expectation value over impulse responses recorded in diverse environments. Therefore, the loss function further updates as:

\begin{equation}
\begin{array}{l}
\underset{\boldsymbol{\delta}}{\operatorname{argmin}} \: \mathbb{E}_{h \sim \mathcal{H}}[\underset{f}{\ell(M F C C(\tilde{\boldsymbol{x}}), t)+\epsilon\|\boldsymbol{\delta}\|}] \\
\text { where } \tilde{\boldsymbol{x}}={\operatorname{Conv_h}}(\boldsymbol{x}+\underset{1k \sim 4k \mathrm{~Hz}}{B P F}(\boldsymbol{\delta}))
\end{array}
\end{equation}
where $\mathcal{H}$ indicates the set of collected impulse responses. The convolution step using impulse response $h$ is shown as $Conv_h$. The last technique the author uses is to add random white Gaussian noise into the generation process. Adding white Gaussian noise to the training process has been proven to help the adversarial example become robust to background noise \cite{hansen1998effective}. Therefore, the final loss function can be described as: 

\begin{equation}
\begin{array}{l}
\underset{\boldsymbol{\delta}}{\operatorname{argmin}} \: \mathbb{E}_{h \sim \mathcal{H}, \boldsymbol{w} \sim \mathcal{N}\left(0, \sigma^{2}\right)}[\underset{f}{\ell}(\operatorname{MFCC}(\tilde{\boldsymbol{x}}), t)+\epsilon\|\boldsymbol{\delta}\|] \\
\text { where } \tilde{\boldsymbol{x}}={\operatorname{Conv_h}}(\boldsymbol{x}+\underset{1k \sim 4k \mathrm{~Hz}}{B P F}(\boldsymbol{\delta}))+\boldsymbol{w}
\end{array}
\end{equation}
where $w$ is a $\mathcal{N}\left(0, \sigma^{2}\right)$ white Gaussian noise. The result shows the attacker reaches a 100\% attack rate by attack JBL CLIP2 and Sony ECM-PCV80 microphone from 0.5 meters away.

\subsubsection{Black box attack}
Until now, all adversarial example generation techniques we introduced are white box attacks, which require the attacker to have complete knowledge of the model architecture and parameters so that the author can compute the gradient of the model and apply the attacks. Recent studies show that black box attacks are also possible in the speech domain \cite{taori2019targeted,biolkova2022neural,mun2022black}. With black box attacks, the attacker does not need information about the speech recognition model so that they can apply the attacks to proprietary systems, such as Google and Amazon APIs.

\begin{figure}[tbp]
        \center{\includegraphics[scale=0.4]
        {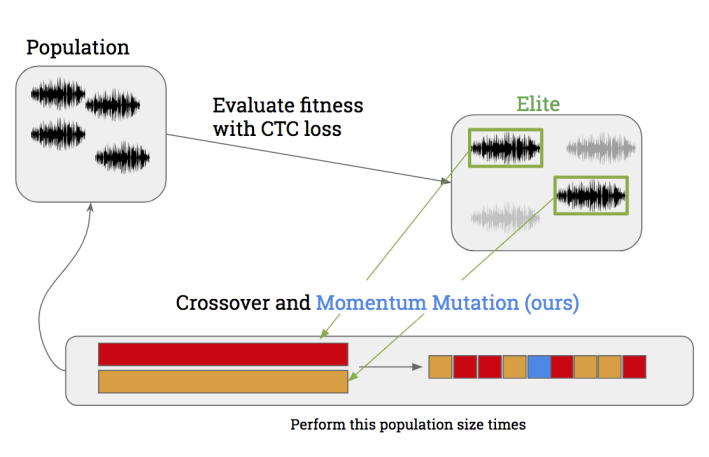}}
        \caption{Genetic algorithm approach from Taori \textit{et al.} \cite{taori2019targeted}.}
        \label{fig:geneticalgorithm}
      \end{figure}

In Taori \textit{et al.} \cite{taori2019targeted}, the author introduces a black box adversarial example method using the CTC loss and Deep Speech. They did not use any gradient information from the model so that this can be treated as a black box attack. The attack scenario contains two stages. In the first stage, the attacker uses a genetic algorithm approach to generate the adversarial audio example, as Fig. \ref{fig:geneticalgorithm} shows. The genetic algorithm approach for adversarial attacks on speech-to-text systems involves iteratively perturbing benign audio samples by applying evolutionary methods like crossover and mutation, using a scoring function based on CTC-Loss to determine the best samples and refining the population over time until the desired target is reached or the maximum number of iterations is completed.

%---------------------------
%% %---------------------------

Given a original input $x$ and target phrase $t$, the algorithm first duplicates the input by the population size the author selects. The author chose 100 as the population size here. Then, on each iteration, the top 10 elite samples with the lowest CTC loss were chosen using a sorting function. These elites then performed crossover and momentum mutation to generate better adversarial examples. The author also added a high pass filter to add noise to the system. After the output of the adversarial sample is close to the target, the attack shifts into the second stage. In the second stage, the author applies a gradient estimation method at 100 random indices of the audio to further fine tune the adversarial example. 

The result is satisfying but not perfect. 35\% accuracy and 89.25\% similarity score is reported for attacking the Deepspeech ASR system. Therefore, even though a black box attack is much more powerful than a white box attack, efforts are still needed in this topic for higher accuracy.

\subsubsection{Universal perturbation attack}
The Universal perturbation attack means the attacker generates a perturbation that can be added to different input audio and cause a misclassify by the ASR system. Universal perturbations have been proven to be effective in image-domain adversarial examples \cite{moosavi2017universal}. Recent studies show that universal perturbations also exist in the audio domain. One thing worth noting is that all the universal perturbation attacks are untargeted, which means they cannot control the resulting transcript. In most cases, the resulting transcription is not a meaningful sentence. Therefore, this type of attack is a limited threat.

In Neekhara \textit{et al.} \cite{neekhara2019universal}, the author applies an untargeted universal perturbation attack on DeepSpeech. The goal of the attack is to find a a quasi-imperceptible universal perturbation $\delta$ that can mis-transcribe most data points sampled from a certain distribution. 
In order to accomplish this task, the author borrowed an idea from the image domain \cite{moosavi2017universal}. The author first went over the data points in the original signals $x$ iteratively and gradually builds the perturbation vector $\delta$. At each iteration, the author finds the minimum perturbation $\Delta \delta_{i}$ that can cause the maximum character error rate (CER). Then they add this perturbation to the desired universal perturbation $\delta$. In order to make the perturbation quasi-imperceptible, the author needs to check  $\|\delta\|_{\infty}<\epsilon$ after each iteration where $\epsilon$ is the maximum allowed $l_{\infty}$ of the perturbation. The result reported a 88.24\% success rate and 1.07 mean CER when maximum allowed $\|\delta\|_{\infty}$ equals 400, where mean $dB_x(\delta) = -30.18$. The success rate and mean CER lower to 72.42\% and 0.82 when the maximum allowed $\|\delta\|_{\infty}$ equals 100, where the mean $dB_x(\delta) = -42.03$. The attack method can also transfer to Wavenet based ASR system with a 63.28\% success rate and 0.6 mean CER when the maximum allowed $\|\delta\|_{\infty}$ equals 400. %This paper bring us a border view of universal perturbation attack on audio.

\section{Defense Approaches}
In this section, we discuss privacy-defending mechanisms to protect user privacy from the previously mentioned attacks. %Different attack mechanisms exploit the vulnerability of voice-controlled systems. More defense mechanisms in the audio domain have formulated defenses against such attacks \cite{mendes2020defending, yang2018characterizing, kwon2019poster}. Recent privacy-defending approaches also borrow ideas from the image domain. For example, in \cite{guo2017countering}, the author uses JPEG compression to increase the robustness of the model. A similar idea has been used in audio papers \cite{das2018adagio} which use AMR and MP3 audio compression techniques as defenses. Another example is \cite{jayashankar2020detecting}, where the author use dropout uncertainty to build a detection network. The idea is also from the image domain \cite{feinman2017detecting}. 
There are two main privacy-defending research directions: detection only and complete defense. In detection only approaches, the defender develops a classifier that determines whether the input from the target user has been attacked or not. For complete defense approaches, the goal is the lower the effectiveness of the attack.

\subsection{Detect Only Defense} \label{sec:detect-only}
A detect only defense is still the most widely used approach for addressing audio privacy attacks. This type of defense is effective to all known attacks and easy to implement. The detect only defense aims to detect when a malicious voice command is inputted to the voice-controlled system. This can help to alarm the user to avoid the incoming adversarial attack.

\subsubsection{Add-on classifier}
The most common detect only defense is to add a classifier to identify whether the input signal is adversarial or not. The defender provides a classifier with a large number of adversarial examples and real speech signals. The classifier then identifies the difference between real and fake speech signals. 

For impersonation attacks, a challenge from The Interspeech conference named Automatic Speaker Verification Spoofing and Countermeasures (ASVspoof) Challenge has provided exciting results. The challenge has been held three times,  in 2015~\cite{wu2015asvspoof}, 2017~\cite{kinnunen2017asvspoof} and 2019~\cite{todisco2019asvspoof}. In the most recent 2019 challenge, both logical access (LA) and physical access (PA) attack scenarios are considered. LA attacks suggest that the attack signal is directly injected into the voice-controlled system. For this scenario, the participants are asked to build a classifier to identify whether the audio is generated by a text-to-speech synthesis (TTS) approach or by voice conversion (VC) technology. For training and validation datasets, 6 known attacks from 2 VC systems and 4 TTS systems are used. For the test dataset, 11 unknown systems with 2 VC, 6 TTS and 3 hybrid TTS-VC systems were chosen to test the generality of the model when they saw unknown data \cite{todisco2019asvspoof}. For the PA scenario, the participants need to build a classification model that distinguishes between human spoken speech and replayed speech. Replayed audio used in the challenge is recorded by 27 different acoustic configurations and 9 different replay configurations. The 27 acoustic configurations include the combination of 3 categories of room sizes, 3 categories of reverberation and 3 categories of talker-to-microphone distance. The 9 replay configurations include 3 attacker-to-talker recording distances and 3 categories of loudspeaker quality. 

Two evaluation metrics are used, where the first one is the tandem detection cost function (t-DCF) \cite{kinnunen2018t} and the second one is the  equal error rate (EER). The EER and t-DCF are both performance measures used to evaluate the trade-off between false acceptances and false rejections. The transitional EER is defined as the point at which the false acceptance rate (FAR) and the false rejection rate (FRR) are equal. 

The t-DCF is defined as:
\begin{equation}
\text{t-DCF} = C_{miss} \times P_{miss} \times \text{EER}_{act} + C_{fa} \times P_{fa} \times \text{EER}_{spoof}
\end{equation}
where $C_{miss}$ and $C_{fa}$ are the costs associated with a missed detection. $P_{miss}$ and $P_{fa}$ are the prior probabilities of a missed detection  and false acceptance, respectively. $\text{EER}_{act}$ is the actual equal error rate (EER) of the system on genuine speech, and $\text{EER}_{spoof}$ is the EER of the system on spoofed speech. The t-DCF can be seen as a weighted combination of the actual EER and the EER on spoofed speech, where the weights are determined by the costs and the prior probabilities. The goal is to minimize the t-DCF, which corresponds to finding a balance between false acceptances and false rejections that is optimal for the given costs and prior probabilities.

\begin{figure}[tbp]
        \center{\includegraphics[scale=0.4]
        {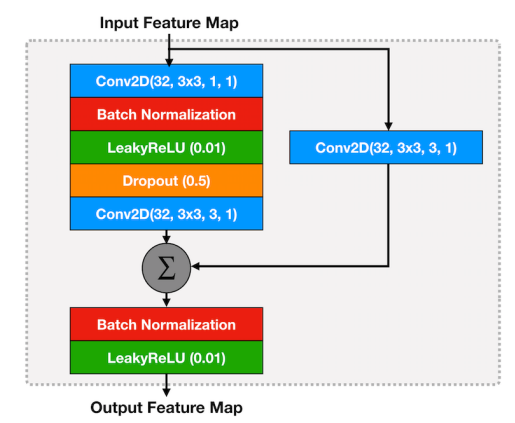}}
        \caption{Detailed structure of residual block used in Alzantot \textit{et al.} \cite{alzantot2019deep}}
        \label{fig:resnet}
      \end{figure}

From the results \cite{todisco2019asvspoof}, the top teams use both neural network based classifiers and an ensemble of classifiers. One representative paper from this challenge is \cite{alzantot2019deep}. In this paper, Alzantot \textit{et al.} use 3 different features: Mel-frequency Cepstral Coefficients (MFCCs), Constant-Q Cepstral Coefficients(CQCCs) and the logarithmic magnitude of the Short-time Fourier transform (STFT). CQCC uses a constant-Q transform and geometrically spaced frequency bins to get  a higher frequency resolution at lower frequencies and higher temporal resolution at higher frequencies. More details about CQCC can be found in Todisco \textit{et al.} \cite{todisco2017constant}. Three different models based on the input were then generated. Three models shared the same structure of a classic classifier from the image domain called ResNet \cite{he2016deep}. A 6 residual block ResNet was chosen. The detailed structure for each residual block can be found in Fig. \ref{fig:resnet}. Two fully connected layers and a softmax layer are connected at the end of the network to produce the probability of whether the audio is fake or not. A fusion mechanism is used to ensemble all MFCC-ResNet, Spec-ResNet and CQCC-ResNet together. A weight is assigned to each model based on its performance on the validation dataset. The result shows that the fusion model reached a 0.1569 t-DCF and 6.02 EER on logical access task, 0.0693 t-DCF and 2.78 EER on physical access task.

%---------------------------
%% %---------------------------

For ultrasonic attacks, in Zhang \textit{et al.} \cite{zhang2017dolphinattack}, the authors develop a support vector machine (SVM) based classifier. The strategy involves analyzing the frequency range from 500 to 1000 Hz, where the attack signal shows differences from both the original signal and the recorded one. To validate the approach, the authors generated 12 voice commands from two different text-to-speech engines, NeoSpeech and Selvy, and obtained both recorded and recovered samples. Using a simple SVM classifier, the approach was able to distinguish recovered audio from recorded ones with 100\% true positive and true negative rates. The results demonstrate the feasibility of using a software-based defense strategy to detect Dolphin attacks \cite{zhang2017dolphinattack}.

In the study by Roy \textit{et al.} [1], the researchers attempt to classify attack signals produced by utilizing three primary characteristics of the created ultrasonic attack voice. Firstly, the attack signal consistently falls below the sub-50Hz band, unlike the human voice. Secondly, there is a strong correlation between the inaudible leaked signal and the ultrasonic attack signal. Thirdly, human voices typically oscillate above or below an amplitude of 0, while the attack signal's amplitude consistently remains above 0. As a result, the amplitude skewness from constructing the ultrasonic attack signal can be employed as an additional feature to discern whether a signal originates from an ultrasonic source. The findings indicate that this detection technique achieves 99\% accuracy in identifying counterfeit voices, as reported by Roy \textit{et al.} \cite{roy2018inaudible}.

For an adversarial example, \cite{samizade2020adversarial} develops a CNN-based classifier to detect C\&W attacks, as we introduced earlier in section \ref{sec:adversarial}. The study presents the design and creation of two separate datasets, A and B, based on white-box and black-box attack methods, respectively. Dataset A is crafted using the C\&W white-box attack method. The audio signals are divided into three categories - short, medium, and long - based on length, and corresponding targets for attacks are composed. Using these categories, 900 adversarial examples and 900 normal examples are generated. Dataset B is created with mutual targeting on Google Speech Command dataset with 10 different commands. The same number of adversarial and normal examples (1800 each) are generated. Both datasets maintain a 16kHz sampling rate. The results are reported under a 95\% confidence interval for all testing scenarios. Matched training and testing conditions achieved over 98\% accuracy, and multi-condition training achieved over 96\% accuracy. These results indicate that the CNN model can effectively learn adversarial perturbation, even with noise present in some normal speech examples. Also, the CNN model performed better in detecting white-box examples than black-box ones.

In D{\"a}ubener \textit{et al.} \cite{daubener2020detecting} and Jayashankar \textit{et al.} \cite{jayashankar2020detecting}, the authors use different uncertainty quantifications (UQ) to detect the adversarial examples. Uncertainty quantification (UQ) is the process of characterizing, modeling and analyzing the uncertainties in a system or model. In D{\"a}ubener \textit{et al.} \cite{daubener2020detecting}, the author uses a feed-forward neural network and three neural networks specifically designed for uncertainty quantification, namely a Bayesian neural network, Monte Carlo dropout, and a deep ensemble and reached 99\% accuracy on detection adversarial examples. 

Jayashankar \textit{et al.} \cite{jayashankar2020detecting} employed dropout uncertainty and a SVM to detect a variety of adversarial examples. By using a defense dropout rate of 0.1 and training the SVM on the first four moments of the character-sequence-based uncertainty distribution, they achieved optimal results. For the C\&W attack, their accuracy was 96.5\%. For the Noise Reduction Robust (NRR) attack, they attained an accuracy of 88.5\%. In the case of the Imperceptible Audio attack, they reached a 92.0\% accuracy, and for the Universal Perturbation, they achieved 100\% accuracy.

The add-on classifier defenses are useful and easy to implement. They did not change any parameters of the original model. However, they also have limitations. First, as the classifiers are built by deep neural networks, they are also vulnerable to adversarial attacks. Also, these classifiers require large number of adversarial data to train. Lastly, their performance still needs to improve for unseen attacks.

\subsubsection{Human motion detection}
Since all the attacks we introduced in section 4 use speakers to play a signal or simply inject noise into the voice-controlled system, another interesting detection only approach for voice-controlled system is that defenders can try to detect whether the signal is from a live human or not.

In Chen \textit{et al.} \cite{chen2017you}, the author uses the magnetic field emitted from the loudspeakers to detect impersonation attacks for voice-controlled system. The defense mechanism tried to detect whether the source of the voice command is from a speaker by using a magnetometer. If the command is from an electronic speaker,  the system will reject the incoming command. The results show that the system reaches 100\% accuracy and 0 EER on detection. However, the system achieved this performance when distances between the sound source and the smartphone were less than or equal to 6 cm. In Lei \textit{et al.} \cite{lei2017insecurity}, the researchers develop a Virtual Security Button (VSButton) that uses WiFi signals to detect indoor human motion. When motion is detected, the voice-controlled system becomes receptive to voice commands. However, there may be instances where a person speaking a voice command does not exhibit detectable motion, resulting in a low true negative rate. The author evaluates the VSButton prototype in three different space settings: a square room, a rectangular room, and a real-world apartment. In the square room experiment, two configurations were tested at  four indoor (A, B, C, D) and six outdoor (A', B', C', D', M', N') locations. In Configuration 1, the Echo Dot laptop (RX) is central in the room, with the WiFi router (TX) at the edge. In Configuration 2, RX and TX sit equidistantly between Locations N' and M', dividing the distance into thirds. The rectangle room is similar to square room Configuration 2 but in a rectangle room with brick walls. The real-world apartment is a 75$m^{2}$ apartment with two bedrooms. The performance is measured by the system's ability to correctly identify three cases: no motion, indoor motion, and outdoor motion. Six volunteers participate in the experiment, performing three different motions - waving a hand, sitting down and standing up, and jumping inside and outside a room, representing weak, medium, and strong human motions, respectively. In the experiment, the receiver (laptop with an Echo dot) sends 50 Internet Control Message Protocol (ICMP) Echo Request messages per second to the transmitter (WiFi router), enabling the continuous collection of channel state information (CSI) for motion detection. In a square room with configuration 1, all indoor motions could be differentiated from no-motion cases and outdoor motions at all locations except location M'. The Mahalanobis distance for the WAVE-HAND motion ranged from 0.191 (location D) to 0.218 (location A) for indoor locations, and from 0.079 (location C') to 0.156 (location M') for outdoor locations. In the same square room with configuration 2, the Mahalanobis distance of each indoor motion was higher than the maximum distance (i.e., 0.241 from JUMP at Location M') of all the outdoor motions. In a rectangular room with brick walls, the minimum Mahalanobis distance among all indoor motions (i.e., 0.147 from WAVE-HAND at Location A) was higher than the maximum distance (i.e., 0.042 from JUMP at Location M') of all outdoor motions. Finally, in the real-world apartment setting, the VSButton was able to differentiate between indoor and outdoor locations with a threshold $t$ set to 0.1. In a 100-minute experiment, the Alexa device was accurately activated by indoor motions and was not activated by outdoor motions. The experimental findings demonstrate that the VSButton is capable of accurately distinguishing between indoor motions and instances of no motion or outdoor motion.

In Feng \textit{et al.} \cite{feng2017continuous}, a new system called VAuth is introduced, which offers continuous authentication for voice-controlled systems. VAuth collects body-surface vibrations from the user and matches them with the voice command captured through microphones on widely-used wearable devices. The researchers implemented a VAuth prototype using a commodity accelerometer and an off-the-shelf Bluetooth transmitter, integrating it with the Google Now system in Android, making it easily extendable to other platforms like Cortana, Siri, or phone banking services. They conducted experiments with 18 participants who issued 30 different voice commands using VAuth in three wearable scenarios: eyeglasses, earbuds, and necklaces. The results showed that VAuth achieved over 97\% detection accuracy and nearly 0 false positives, indicating successful command authentication. It worked effectively across different accents, mobility patterns (still vs. jogging), and languages (Arabic, Chinese, English, Korean, Persian). VSAuth successfully blocked unauthenticated voice commands replayed by an attacker or impersonated by other users and it incurred minimal latency (average of 300ms) and energy overhead (requiring recharging only once a week). However, VAuth's reliance on users wearing these devices can be inconvenient for everyday use.

Exploring human motion detection as a potential avenue for detecting adversarial attacks holds significant promise for future studies. However, recent work has been limited by hardware constraints and has not yet achieved satisfactory real-world protection. As such, it is necessary to conduct additional research that utilizes alternative features for detecting human motion in order to address this limitation.

\subsection{Complete Defense}
Compared to detect only defense techniques, complete defenses are more powerful. This type of defense uses different techniques to modify the original model setup and help the system become robust to certain types of adversarial attacks. When the attack command reaches the voice-command system, the target network could achieve its original goal and provide the correct output. However, complete defenses also have their limitations. When unknown attacks arrive, the modified system may fail. 

There are two advantages of this type of defense. Firstly, it can be widely used in different types of attacks \cite{rajaratnam2018noise, lei2017insecurity, wang2017feature}. Secondly, this type of defense is easy to implement, as the defense does not require changes to be made or the system to be retrained. Detection only defenses also have their weaknesses, since they can only detect if the attack occurs but they cannot resolve the problem. The complete defense approach addresses this, by lowering the effectiveness of the attack. Complete defense approaches are normally used to defend against adversarial attacks. This type of defense has been proven to be useful in the image domain \cite{zantedeschi2017efficient, gao2017deepcloak, akhtar2018defense, lee2017generative}. More audio privacy-defending approaches also aim to strengthen the deep learning model. The complete defense approaches also have their weakness. Firstly, unlike detection only defense mechanisms which usually have 90\% or high accuracy, the complete defense can only lower the effectiveness of the adversarial examples in a certain extent. Secondly, most complete defense algorithms only work for known attacks. When unknown attacks appear, the modified model may crash.

\subsubsection{Hardware-based defense}
In Zhang \textit{et al.} \cite{zhang2017dolphinattack}, the author provided two hardware based defenses to avoid ultrasonic attacks. The first and straight forward way is to enhance the microphone. Most MEMS microphones now still allow signals in ultrasonic range (>20kHZ) \footnote{\url{http://www.mouser.com/ds/2/720/DS37-1.01\%20AKU143\%20Datasheet-552974.pdf}} \footnote{\url{https://www.mouser.com/datasheet/2/720/PB24-1.0\%20-\%20AKU242\%20Product\%20Brief-770082.pdf}}. Therefore, the author suggested these type of microphones should be enhanced and apply filters to eliminate any signals within the ultrasonic range. The second defense mechanism the author proposed is to add a module before the low pass filter to detect the ultrasonic signal and cancel the baseband of it. These two methods can efficiently defend against weak ultrasonic attacks. However, for stronger ultrasonic attacks such as \cite{roy2018inaudible}, these type of defense may not work.

\subsubsection{Increasing security level}
In Petracca \textit{et al.} \cite{petracca2015audroid}, the author proposed a security level increasing method to prevent adversarial attacks on audio channels in mobile devices. The authors design and implement AuDroid, an extension to the SELinux reference monitor integrated into the Android OS. AuDroid enforces lattice policies on the dynamic use of system audio resources and gathers input from system apps and services to evaluate options for resolving unsafe communication channels. The system is specifically integrated into the Android Media Server, controlling access to the microphone and speaker to protect system apps from third-party apps and external attackers. AuDroid is evaluated on six types of attack scenarios and on 17 widely-used apps that utilize audio. The results show that AuDroid effectively prevents exploits without impairing the normal functionality of system apps and services, with defense times of less than 4 µs for the speaker and less than 25 µs for the microphone, resulting in insignificant overhead during app usage. This type of defense allows the system to defend the operating system attack by using the built-in speakers \cite{diao2014your}. However, the limitation of this defense is that it is not robust to other types of attacks, such as adversarial examples.

\subsubsection{Modify input data}
Modify input data for adversarial training has proven its effectiveness in the image domain \cite{goodfellow2014explaining, sankaranarayanan2018regularizing}. This defense uses adversarial training to regularize the network, reduce over-fitting and add robustness to the network. 

In Sun \textit{et al.} \cite{sun2018training}, the authors incorporate adversarial examples directly into the training dataset to train the ASR system. During the training process, the Fast Gradient Sign Method (FGSM) is employed to generate adversarial examples as training data. The underlying concept is to create adversarial examples that maximize the loss function. The perturbation can then be calculated as:

\begin{equation}
\boldsymbol{\delta}_{F G S M}=\epsilon \operatorname{sign}\left(\nabla_{\boldsymbol{x}} J(\boldsymbol{\theta}, \boldsymbol{x}, t)\right)
\end{equation}
where $\epsilon$ is a constant that controls the amplitude of the perturbation , $x$ represents the input, $t$ denotes the target and $\theta$ signifies the model's parameters. $J(\boldsymbol{\theta}, \boldsymbol{x}, t)$ is the average cross-entropy loss function. The $sign$ operation in FGSM yields the direction of the gradient, either positive or negative, instead of its actual value, facilitating better control over the quantity of perturbation introduced.The perturbation is dynamically generated during each iteration, which helps to improve the robustness of the ASR system. By setting $\epsilon$ to 0.3, the results demonstrate an average 14.1\% WER reduction. While this method effectively enhances robustness to defend against known attacks, it has its limitations. When an unknown adversarial example is inputted into the ASR system, it may still be susceptible to deception.

\begin{figure}[tbp]
        \center{\includegraphics[scale=0.37]
        {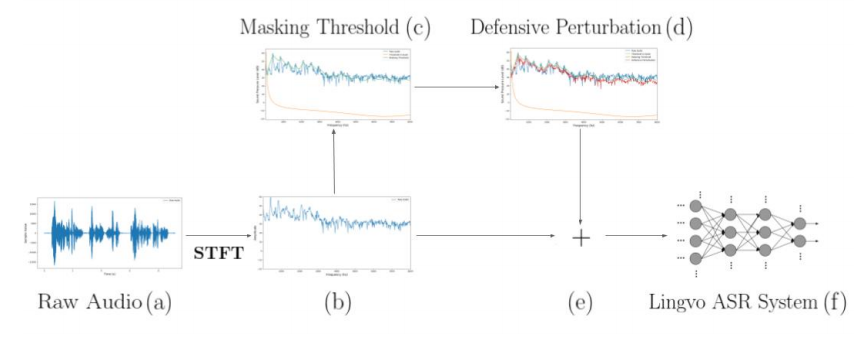}}
        \caption{Detailed structure of the defense mechanism from Mendes \textit{et al.} \cite{mendes2020defending}}
        \label{fig:addnoise}
      \end{figure}

In Mendes \textit{et al.} \cite{mendes2020defending}, the authors proposed a complete defense method to protect speech from adversarial examples. The overview of this defense can be found in Fig. \ref{fig:addnoise}. The raw audio signal is first transformed into the time-frequency domain using a Short-time Fourier transform (STFT). Then the frequency masking threshold $\theta_{x}$ is calculated by using the method from Qin \textit{et al.} \cite{qin2019imperceptible}. A corresponding defensive perturbation is calculated by $\delta_{D}=\max (0, \mathcal{N}(\mu, \sigma))$ where $\mu:=3 k \times \theta_{x}$ and $\sigma:=k \times \theta_{x}$. $k$ here denotes the proportionality. The last step of this defense is to add the perturbation to the original input and feed into ASR model.

%---------------------------
%% %---------------------------

\subsubsection{Modify the network}
Besides modifying the input and using data augmentation techniques to make the model more robust to adversarial examples, we can also directly modify the network. In Yang \textit{et al.} \cite{yang2018characterizing}, the author provided an additional term to the network based on the temporal dependency between a real speech and an adversarial example. For a given audio, the author first selected the prefix of length $p$ and provided x to a ASR system to generate the transcript  $S_{p}$. Then The complete audio signal was inputted into the ASR system, and the transcribed result prefix of length $p$ was selected as  $S_{\{\text {whole }, p\}}$. Due to the temporal dependency, $S_{p}$ and $S_{\{\text {whole }, p\}}$ should have consistent results. However, if the speech has been attacked with an added perturbation, they will not produce the same result. The study evaluated the proposed Temporal Dependency (TD) detection method on speech-to-text attacks - Commander Song and Optimization based attack(Opt) \ref{sec:adversarial}. For the Commander Song attack, the TD method with p=1/2 successfully detected all generated adversarial samples. In the Opt attack, the TD method achieved an AUC score of 0.936 on Common Voice and 0.93 on LIBRIS when using WER as the detection metric. When p=4/5 and using CER, the AUC score reached 0.969, indicating that the TD-based method is promising in distinguishing adversarial instances. The results suggest that the TD-based method is an easy-to-implement and effective approach for characterizing adversarial audio attacks.

\subsubsection{Audio compression}
Data compression has emerged as a popular method of adversarial defense in the image domain \cite{dziugaite2016study}, and similar ideas have been implemented in the audio domain. In Das \textit{et al.} \cite{das2018adagio}, the study explores the use of compression techniques, such as Adaptive Multi-Rate audio codec(AMR) and MP3 compression, to mitigate adversarial perturbations in the audio domain. The researchers tested these techniques on adversarially manipulated audio samples and evaluated their effectiveness in defending an ASR model. They created targeted adversarial instances from the first 100 test samples of the Mozilla Common Voice dataset and preprocessed them using the compression techniques.  The proposed system significantly reduces the attack success rate from 92.5\% to 0\%. The results indicate that the word error rate (WER) of the ASR system without any defense increased from 0.369 to 1.287. The WER slightly increased to 0.666 from 0.488 when using AMR compression, and to 0.78 from 0.4 when using MP3 compression. Likewise, in Andronic \textit{et al.} \cite{andronic2020mp3}, the authors use MP3 compression to eliminate adversarial noise for the ASR system, resulting in a 21.31\% reduction in the relative character error rate of adversarial examples and MP3-compressed adversarial examples. These compression techniques, based on psychoacoustic principles, were found to be effective in removing adversarial components from the audio that are imperceptible to humans but confuse the model.

\subsubsection{Adversarial Defense}
The tactics used for attacking others can also serve to protect them. Adversarial examples are increasingly being employed by researchers to safeguard users' privacy from voice-controlled assistants. For instance, Liu \textit{et al.} \cite{liu2021defending} have devised "MyBabble," which uses the user's own voice to generate personalized noise that thwarts speech hijacking by voice-controlled assistants. Similarly, Liu \textit{et al.} \cite{liu2022preventing} have implemented an end-to-end approach that produces utterance-specific perturbations that obscure a set of words considered sensitive. In Xu \textit{et al.} \cite{xu2018hasp}, the authors utilize adversarial noise based on MFCCs to defend users against malicious speech recognition (ASR) systems, thereby raising the system's word error rate. Additionally, in Chen \textit{et al.} \cite{chen2020wearable}, the authors created a wearable microphone jammer that emits ultrasonic sounds to protect people's conversations from being overheard by voice-controlled devices.

\section{Discussion}
In the previous sections, we conducted a comprehensive list of privacy-attacking and privacy-defending techniques. We introduced the theory behind each mechanism. We also talked about the limitations and advantages about certain techniques. In this section, we discussed our key findings and make recommendations for future research. \\

\noindent{\bf Combination of attacks may become a big threat.} Until now, we define each attack based on when it occurs in the voice-controlled system pipeline. However, different attacks may combine these attacks, resulting in stronger attacks. For example, an operating system attack can be combined with an adversarial example so that the malware causes the built-in microphone to play the inaudible adversarial example. In this case, this type of attack is much harder to detect and defend. \\

\noindent{\bf Reasons for adversarial vulnerability need more investigation.} The literature shows that all kinds of deep learning networks are vulnerable to certain attacks. In our previous explanation, non-linearity is the key feature of the existence for those attacks. However, more investigations are needed for exploring the common feature of these attacks. Current complete defense methods still cannot fully defend certain known attacks. More investigations are needed to find new features for those attacks and build more robust ASR defense models. Also, many counter measurements are based on DNNs. Theses defenses are also vulnerable to adversarial attacks. A more general and stable defense may be needed. \\

\section{Conclusion}
Modern voice-controlled systems are vulnerable to privacy attacks. In this paper, we proposed a categorization for privacy-attacking and privacy defending mechanisms. We carefully introduced each attacking and defending technique with their threat model. These privacy attacks can happen in each stage of the voice-controlled system and they pose real-world threats to our daily life. Privacy-defending techniques can make the system more robust but still cannot completely solve the problem. More studies are needed on voice-controlled systems to ensure privacy is preserved for users.

\bibliographystyle{ACM-Reference-Format}
\bibliography{ref}%ccs-sample}
\end{document}